\documentclass[%
reprint,
superscriptaddress,
 amsmath,amssymb,
 aps,
prb,
]{revtex4-1}

\usepackage {graphicx,epsfig,graphics,color}
\usepackage{dcolumn}
\usepackage{bm}
\usepackage{hyperref}
\usepackage{sidecap}
\usepackage{float}

\begin{document}
\title{Emergent nematicity and intrinsic vs. extrinsic electronic scattering processes in the kagome metal CsV$_3$Sb$_5$}

\author{Dirk Wulferding}
\email[]{Corresponding author: dirwulfe@snu.ac.kr}
\altaffiliation{Contributed equally to this work.}
\affiliation{Center for Correlated Electron Systems, Institute for Basic Science, Seoul 08826, Republic of Korea}
\affiliation{Department of Physics and Astronomy, Seoul National University, Seoul 08826, Republic of Korea}

\author{Seungyeol Lee}
\altaffiliation{Contributed equally to this work.}
\affiliation{Department of Physics, Chung-Ang University, Seoul 06974, Republic of Korea}

\author{Youngsu Choi}
\affiliation{Department of Physics, Sungkyunkwan University, Suwon 16419, Republic of Korea}

\author{Qiangwei Yin}
\affiliation{Department of Physics and Beijing Key Laboratory of Opto-electronic Functional Materials \& Micro-nano Devices, Renmin University of China, Beijing 100872, China}

\author{Zhijun Tu}
\affiliation{Department of Physics and Beijing Key Laboratory of Opto-electronic Functional Materials \& Micro-nano Devices, Renmin University of China, Beijing 100872, China}

\author{Chunsheng Gong}
\affiliation{Department of Physics and Beijing Key Laboratory of Opto-electronic Functional Materials \& Micro-nano Devices, Renmin University of China, Beijing 100872, China}

\author{Hechang Lei}
\email[]{Corresponding author: hlei@ruc.edu.cn}
\affiliation{Department of Physics and Beijing Key Laboratory of Opto-electronic Functional Materials \& Micro-nano Devices, Renmin University of China, Beijing 100872, China}

\author{Saqlain Yousuf}
\affiliation{Department of Physics, Sungkyunkwan University, Suwon 16419, Republic of Korea}

\author{Jaegu Song}
\affiliation{Department of Physics, Sungkyunkwan University, Suwon 16419, Republic of Korea}

\author{Hanoh Lee}
\affiliation{Department of Physics, Sungkyunkwan University, Suwon 16419, Republic of Korea}

\author{Tuson Park}
\affiliation{Department of Physics, Sungkyunkwan University, Suwon 16419, Republic of Korea}

\author{Kwang-Yong Choi}
\email[]{Corresponding author: choisky99@skku.edu}
\affiliation{Department of Physics, Sungkyunkwan University, Suwon 16419, Republic of Korea}

\date{\today}

\begin{abstract}
Fermi surface fluctuations and lattice instabilities in the 2D metallic kagome superconductor CsV$_3$Sb$_5$ are elucidated via polarization-resolved Raman spectroscopy. The presence of a weak electronic continuum in high-quality samples marks the cross-over into the charge-density-wave (CDW) ordered phase, while impurity-rich samples promote strong defect-induced electronic scattering processes that affect the coherence of the CDW phase. CDW-induced phonon anomalies appear below $T_{\mathrm{CDW}}$, with emergent $C2$ symmetry for one of the CDW amplitude modes, alluding to nematicity. In conjunction with symmetry-breaking lattice distortions, a kink-like hardening of the A$_{1g}$ phonon energy at $T_{\mathrm{CDW}}$ signifies a concerted interplay of electronic correlations and electron-phonon coupling in the exotic CDW order.

\end{abstract}

\maketitle

\section{Introduction}

Two-dimensional kagome metals are at the forefront of correlated topological physics as their electronic structure entails flat bands, van Hove singularities, and Dirac cones. The sought-after many-body states in kagome materials include topological superconductivity, unconventional charge density wave (CDW), fractional quantum Hall, and Majorana fermions~\cite{ko-09,guo-09,ruegg-11,wang-13,kiesel-13,mazin-14,nakatsuji-15,ye-18,yin-20}.

The recently discovered kagome metals $A$V$_3$Sb$_5$ ($A$=K, Rb, and Cs; $P6/mmm$ space group) have opened a new avenue toward establishing the relationship between CDW, anomalous Hall effect (AHE), and superconductivity (SC)~\cite{ortiz-19}. In this V-based kagome family, an alkali metal layer and a V$_3$Sb$_5$ layer are alternately stacked, forming quasi-2D kagome layers of the V ions. CsV$_3$Sb$_5$ features $\mathbb{Z}_2$ band topology and a two-stage symmetry breaking: a CDW transition at $T_{\mathrm{CDW}}\sim 95$~K with a subsequent superconducting transition at $T_c \sim 2.7$~K~\cite{ortiz-19,ortiz-20,yang-20,ortiz-21, qwyin-21}. The CDW order with inverse star-of-David pattern involves $2\times 2$ and $1\times 4$ charge orders~\cite{Jiang-21,tan-21}. The nesting of the Fermi surface assisted by the M-point van Hove singularity may provide a primary cause of the CDW transition~\cite{Jiang-21}. However, recent ARPES measurements unveil that the alternating stacking of Star-of-David and tri-hexagonal distortions brings about three-dimensional $2\times2\times2$ or $2\times2\times4$ CDW orders and their coexistence~\cite{hu-22}. Strikingly, a large AHE concurs with the CDW order despite the lack of long-range magnetic ordering~\cite{ortiz-19, yang-20, Jiang-21, yu-21a, li-21, kenney-21}. Orbital currents have been invoked as the origin of spontaneous time-reversal symmetry breaking~\cite{tan-21, feng-21, denner-21, park-21, mielke-21, Lyu-21, Hli-21}. Furthermore, contradictory views exist on the SC pairing mechanism and symmetry: unconventional nodal superconductivity~\cite{zhao-21,liang-21,chen-21} vs. an $s$-wave superconductor~\cite{duan-21,mu-21} or multiband superconductivity with sign-preserving order parameter~\cite{xu-21,nakayama-21,gupta-21}.

To understand the link between the CDW instability and superconductivity, it is pivotal to elucidate both Fermi surface and lattice instabilities below the CDW state. In this work, we employ polarization-resolved Raman spectroscopy to probe low-energy electronic fluctuations and phonon anomalies in CsV$_3$Sb$_5$. By comparing samples of varying quality, we disentangle defect-induced scattering contributions from intrinsic scattering processes. We uncover CDW-associated lattice distortions and persistent dynamical scattering down to $T = 5$ K. The angular-dependent Raman data reveal the emergence of a subtle twofold anisotropy at $T = 5$ K of the E$_{2g}$ amplitude mode of CDW order. In addition to nematicity, the change of phonon self-energy below $T_{\mathrm{CDW}}$ highlights the significance of electron-phonon coupling.

\section{Experimental Details}

Two single crystals of CsV$_3$Sb$_5$ with slightly different sample qualities were employed in the present study, in the following denoted as ``sample A'' and ``sample B''. Sample A was obtained via a self flux method from Cs (Alfa, 99.9 \%), V (Kojundo, 99.7 \%), and Sb (Alfa, 5N). The stoichiometric raw materials were inserted into an alumina crucible with excess Sb and sealed in a quartz tube under partial Ar atmosphere. The sealed ampoule then was heated to 1000$^{\circ}$C and stayed for 12 h. The molten mixture was cooled slowly to 670$^{\circ}$C at 3$^{\circ}$C/h, where it was centrifuged to remove the extra Sb-flux, resulting in cm-sized plate-like single crystals. Sample B was grown from Cs ingot (purity 99.9 \%), V powder (purity 99.9 \%) and Sb grains (purity 99.999 \%) using the self-flux method, similar to the growth of RbV$_3$Sb$_5$~\cite{qwyin-21}. The eutectic mixture of CsSb and CsSb$_2$ is mixed with VSb$_2$ to form a composition with approximately 50 \% Cs$_x$Sb$_y$ and 50 \% VSb$_2$. The mixture was placed into an alumina crucible and sealed in a quartz ampoule under a partial argon atmosphere. The sealed quartz ampoule was heated to 1273 K for 12 h and soaked there for 24 h. It was subsequently cooled down to 1173 K at 50 K/h and further to 923 K at a slower rate. Finally, the ampoule was removed from the furnace and decanted with a centrifuge to separate CsV$_3$Sb$_5$ single crystals from the flux. The obtained crystals have a typical size of $2\times 2 \times 0.02$~mm$^3$. To avoid degradation of the sample surface, a freshly cleaved single crystal was mounted onto the cold-finger of a helium-flow cryostat inside an Ar atmosphere glove box. Raman spectroscopic measurements were carried out using a $\lambda = 561$ nm laser (Oxxius LCX) and a Princeton Instruments TriVista spectrometer equipped with a volume Bragg grating notch filter (Optigrate; low-energy cut-off: 6 cm$^{-1}$) and a liquid-Nitrogen cooled CCD (PyLoN eXcellon). The incident laser was focused onto the sample with a 40$\times$ microscope objective resulting in a spot diameter of about 5 $\mu$m. Its power was kept at a power level of 0.1 mW to minimize heating effects. Polarization-resolved angular-dependent experiments were carried out using a rotatable superachromatic $\lambda$/2 wave plate positioned above the microscope objective.

\section{Results}

\begin{figure}
\label{figure1}
\centering
\includegraphics[width=8cm]{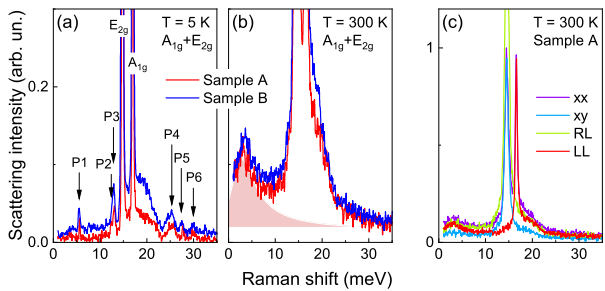}
\caption{Comparison of Raman spectra obtained in A$_{1g}$+E$_{2g}$ symmetry for two samples of different quality at (a) $T = 5$ K and (b) at $T = 300$ K. CDW-induced modes are labeled P1--P6. The shaded area marks the quasi-elastic peak for sample A. (c) Polarization-resolved Raman spectra measured at $T = 300$ K for sample A.}
\end{figure}

\subsection{Electronic Raman scattering and phonons.}
In Figs. 1(a) and 1(b) we show representative Raman spectra obtained at $T = 5$~K and 300~K for two specimens, samples A (red line) and B (blue line). Figure 1(c) compares Raman spectra of sample A obtained at $T = 300$~K in different polarization configurations. Based on these, we can assign a phonon at around 15 meV to E$_{2g}$ symmetry, and a phonon at around 17 meV to A$_{1g}$ symmetry, in accordance with other recent Raman studies~\cite{wu-22, liu-22}. While spectra of samples A and B closely resemble each other, there are subtle differences, hinting at a difference in sample quality. At base temperature [see Fig. 1(a)], sample A exhibits no low-energy electronic response (i.e., a flat background at low energies), with a minor, broad excitation centered around 17 meV. The low-energy background scattering for sample B, on the other hand, is slightly raised, and in particular, the mid-energy continuum around 17 meV is significantly enhanced. At $T>T_{\mathrm{CDW}}$, both the rather symmetric maximum centered around 17 meV and the low-energy quasielastic peak (QEP) increase in intensity [see Fig. 1(b)], where sample B again exhibits slightly larger scattering intensities of both contributions. There are different possible scattering mechanisms from which these contributions may originate. One possibility is that the QEP scattering pertains to a long-wavelength dynamical electronic response, and the mid-energy continuum is related to an interband transition between van-Hove singularity bands near the M point. On the other hand, the observed sample dependence might suggest a phonon density of states for the mid-energy continuum. This latter assumption is further supported by its rather weak polarization dependence [see Fig. 1(c)]. The systemic suppression of the QEP detected in the high-quality sample A through $T_{\mathrm{CDW}}$ suggests that it is related to intrinsic electronic scattering. More specifically, the gradual vanishing of the QEP scattering below $T_{\mathrm{CDW}}$ reflects the partial depletion of the Fermi surface induced by the CDW transition. In contrast, the QEP of sample B shows a persistent, residual scattering intensity at the lowest measured temperature, implying that  additional scattering channels, involving mainly extrinsic defect-induced scattering processes, dominate over the intrinsic ones.

\subsection{Thermal evolution and CDW-induced phonon anomaly.}

\begin{figure*}
\label{figure2}
\centering
\includegraphics[width=18cm]{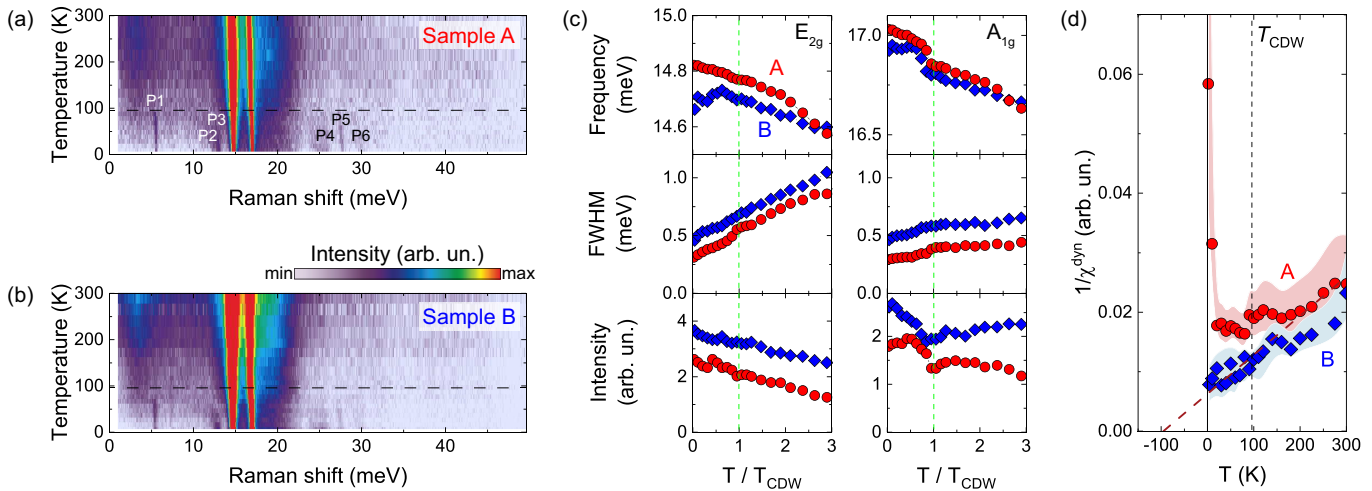}
\caption{(a) Temperature-dependent Raman spectra of Sample A. (b) Temperature-dependent Raman spectra of Sample B. Both panels plot as-measured Raman intensity obtained in the A$_{1g}$+E$_{2g}$ channel. (c) Temperature-dependence of phonon parameters extracted from Sample A (red circles) and Sample B (blue diamonds). Dashed lines denote $T_{\mathrm{CDW}}$. (d) Inverse dynamic Raman susceptibility as a function of temperature for sample A and sample B. The dashed dark red line is a Curie-Weiss fit to the high temperature data of sample A. Shaded areas denote error bars.}
\end{figure*}

To elucidate phonon anomalies associated with the CDW transition, we analyze the temperature-dependent phonon behavior in detail. Recent numerical \textit{ab initio} calculations highlighted a possible symmetry reduction from $P6/mmm$ to $C2/m$ together with phonon softening at the M, L, and X points of the Brillouin zone to facilitate CDW formation~\cite{ptok-21}.

Figures 2(a) and 2(b) present color contour plots of the temperature-dependent Raman data measured for samples A and B, respectively. For both samples the two dominating phonons at 15 meV and 17 meV evidence subtle changes around the CDW transition temperature. A second feature commonly observed is the low-energy QEP, which grows in intensity as the temperature is raised. Here we notice a stronger response in sample B compared to sample A, and the QEP can be clearly traced into the CDW phase, while for sample A it only clearly emerges above $T_{\mathrm{CDW}}$. Besides these two phonons and the low-energy background, new CDW-induced modes appear below $T_{\mathrm{CDW}}$, and at base temperature these modes settle at 5.6 meV (P1), 12.5 meV (P2), 13 meV (P3), 25.4 meV (P4), 27.7 meV (P5), and 30 meV (P6). While there is generally a good agreement between the thermal evolution observed in samples A and B, we note important differences: CDW-induced modes persist up to $T_{\mathrm{CDW}}$ in sample A, whereas in sample B they already diminish about 10-20 K below the transition temperature. Additionally, their linewidths appear much sharper in sample A compared to those in sample B. Both observations are indicative of the sensitivity of the CDW phase to disorder and impurities. In sample B, the CDW mode located around 12.5 meV, and thereby close to the A$_{1g}$ phonon, appears to develop on the left shoulder of the defect-enhanced electronic continuum below $T^*\sim 50$~K. This mode, involving Sb vibrations, dramatically loses in intensity and broadens significantly well below $T_{\mathrm{CDW}}$. Its rapid repression and damping indicates its strong susceptibility to the competition of the two distinct CDW orders and Sb disorders. Concomitantly, the E$_{2g}$ phonon energy of sample B shows a kink around $T^*$. These two characteristic temperatures $T^*$ and $T_{\mathrm{CDW}}$ of sample B imply the presence of two types of charge modulations deep inside the CDW state. Indeed, STM and $^{51}$V NMR studies of CsV$_3$Sb$_5$ reveal another charge modulation at 40~K, which is interpreted in terms of a $4a_0$ charge order pattern or a roton pair density wave~\cite{chen-21, luo-21}. However, this cross-over in sample B may also be a result of stacking disorder, as a clear $T^*$ is absent in the high quality sample A.

In Fig. 2(c), we present the phonon parameters (energy, linewidth, and intensity) as a function of temperature for the two phonons measured in sample A (red circles) and sample B (blue diamonds). Clear evidence for coupling between electronic degrees of freedom and in-plane phonons can be deduced from the temperature dependence of the phonon self-energy: a softening can be seen in the energy of the E$_{2g}$ phonon. In contrast, the A$_{1g}$ phonon follows conventional anharmonic behavior only down to $T_{\mathrm{CDW}}$. At lower temperatures, it pronouncedly hardens, together with a discontinuous increase in intensity. In addition, for sample A both phonons show a clear drop in linewidth at the transition temperature. The deviation of these phonon energies from anharmonic softening is related to the CDW gap, as a phonon softening occurs for $\omega_{\mathrm{phonon}} < \Delta_{\mathrm{CDW}}$, and a hardening for $\omega_{\mathrm{phonon}} > \Delta_{\mathrm{CDW}}$~\cite{zeyher-90}. Here, $\Delta_{\mathrm{CDW}} = 15-17$ meV should be regarded as an effective CDW gap probed by Raman spectroscopy. Overall, the larger linewidth in sample B compared to sample A directly underlines its slightly inferior sample quality.

Additional discrepancies between samples A and B become evident when analyzing the dynamic Raman susceptibility $\chi^{\mathrm{dyn}}(T)$ [see Fig. 2(d)]. This quantity allows us to access dynamic electronic fluctuations for a given symmetry channel, thereby we may quantify electronic correlations~\cite{gallais-16}. $\chi^{\mathrm{dyn}}(T)$ is approximated by integrating $\chi''(\omega)/ \omega$ from 1~meV up to 12~meV. Here, $\chi''(\omega)$ is the Bose-corrected as-measured Raman intensity. For sample A the high-temperature data, up to about 150 K, can be described by a Curie-Weiss law $\chi^{\mathrm{dyn}}(T)\propto \frac{1}{T-T_{\mathrm{CW}}}$ with the Curie-Weiss temperature $T_{\mathrm{CW}} \approx -95$ K, i.e., $\mid T_{\mathrm{CW}}\mid \approx T_{\mathrm{CDW}}$ (see the dark red dashed line). For sample B it yields $T_{\mathrm{CW}} \approx -160$ K. Below 150 K the dynamic susceptibility for sample A starts to deviate from its Curie-Weiss behavior as electron-electron correlations increase upon approaching $T_{\mathrm{CDW}}$. A small discontinuous jump marks the transition into the CDW-ordered phase in sample A as $\chi^{\mathrm{dyn}}(T)$ begins to strongly deviate from a Curie-Weiss behavior. This suggests that here the dynamical Raman response traces Fermi surface instabilities. Note that the divergent behavior of the inverse susceptibility reflects the vanishing low-energy scattering intensity (and thereby the vanishing dynamic susceptibility) as $T \to 0$. For sample B we obtain a very different temperature dependence, with the three major observations: (i) the lacking divergence of $\chi^{\mathrm{dyn}}(T)$ at $T_{\mathrm{CDW}}$, (ii) the Curie-like behavior of $\chi^{\mathrm{dyn}}(T)$ down to the SC state, and (iii) a negative $T_{\mathrm{CW}}$ value at high temperatures of about -160 K that does not reflect any other intrinsic temperature scale of CsV$_3$Sb$_5$. All these observations support the notion that the dynamical Raman response in sample B is dictated largely by extrinsic scattering mechanisms other than the CDW order, e.g., electronic Raman scattering enabled and enhanced by defects, impurities, and stacking faults. In contrast to the superior sample A, these extrinsic scattering processes mask any intrinsic electronic Raman scattering present in sample B.

\subsection{Symmetry breaking and nematicity.}

\begin{figure*}
\label{figure3}
\centering
\includegraphics[width=12cm]{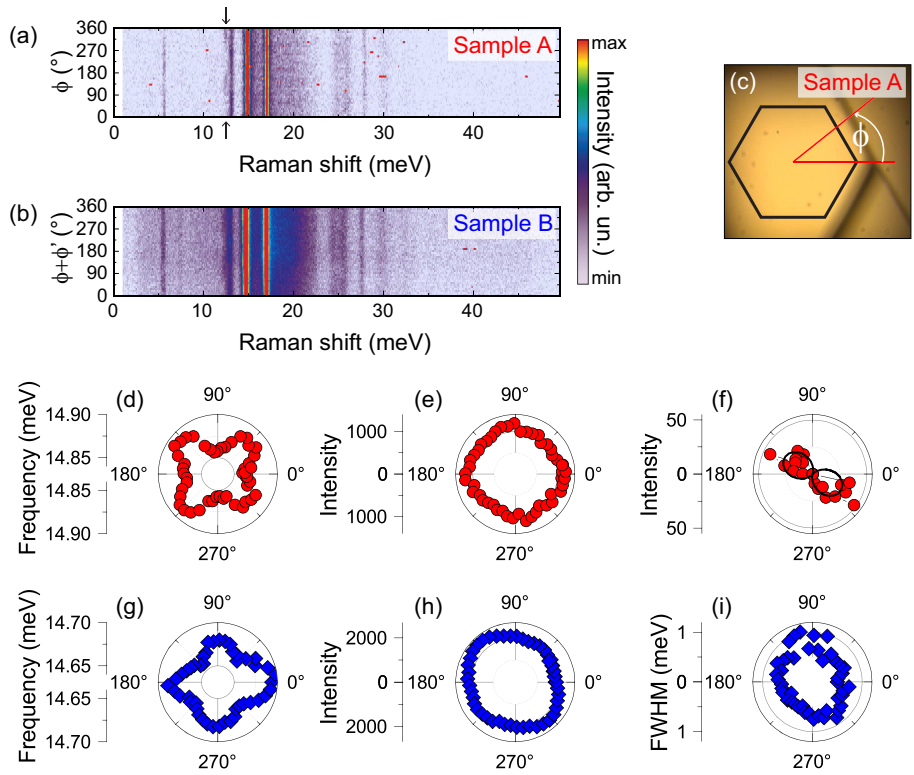}
\caption{Raman scattering intensity measured in parallel polarization at $T=5$ K with the light polarization rotated in-plane from 0$^{\circ}$ to 360$^{\circ}$ on (a) sample A, and (b) sample B. (c) A sketch of the orientation of sample A with respect to the polarization angle $\phi$. E$_{2g}$ phonon frequency (d) and intensity (e) of sample A as a function of polarization angle $\phi$. (f) Two-fold polarization-dependence of the CDW mode intensity detected in sample A at 12.5 meV and outlined by a fit (black line). Polarization-dependence of the E$_{2g}$ phonon frequency (g) and intensity (h) of sample B. (i) Linewidth (FWHM) of the merged CDW excitations centered around 13 meV in sample B.}
\end{figure*}

Further evidence for instabilities and related symmetry breaking can be obtained from highly-resolved polarization-dependent Raman measurements. In Figs. 3(a) and 3(b) we show the as-measured Raman intensity at $T=5$ K for both samples while rotating the polarization of the laser light within the crystallographic $ab$-plane. Here, the light polarization of the incident light and that of the scattered light are kept parallel, thereby probing the A$_{1g}$+E$_{2g}$ channel. Fig. 3(c) sketches the polarization of the light with respect to the underlying crystallographic axes, as deduced from a characteristic as-grown edge on sample A's surface shown in the microscope image. For both samples A and B, no dramatic polarization dependence is observed and the signal is almost isotropic. There are however subtle polarization dependencies detected. Most remarkably, the E$_{2g}$ phonon at 15 meV evidences a clearly four-fold rotational modulation of its frequency, with a modulation amplitude of about 50 $\mu$eV [Fig. 3(d)]. Similarly, its intensity follows a slight four-fold distorted pattern [Fig. 3(e)]. The linewidth on the other hand (not shown here) appears constant as a function of polarization. A frequency modulation of this two-fold degenerate E$_{2g}$ mode suggests a subtle lifting of degeneracy as a consequence of CDW order. This is further supported by an analysis of the CDW mode at 12.5 meV, marked by black arrows in panel (a). In clear contrast to other CDW-induced modes, the intensity of this excitation follows a two-fold symmetry, shown in Fig. 3(f). Its lobes reach maximum intensity at about 90$^{\circ}$ off one of the as-grown edges, i.e., orthogonal to the crystallographic $a$ or $b$ axis. For sample B, similar four-fold rotational anisotropies are found in the frequency and the intensity of the E$_{2g}$ phonon [shown in Figs. 3(g)-(h)]. These polarization plots appear rotated by about 45$^{\circ}$ from those of sample A. The shape of sample B lacked any characteristic morphology, it was therefore not possible to connect the polarization angle to the in-plane sample orientation. Consequently, we plot the Raman spectra in Fig. 3(b) as a function of polarization angle $\phi+\phi'$, where $\phi'$ is undefined. Yet, based on the rotational similarities between both samples, it is reasonable to assume a 45$^{\circ}$ in-plane offset between samples A and B, and hence $\phi'\approx45^{\circ}$. In contrast to sample A, the CDW modes appear with significantly increased linewidths and the two excitations at 12.5 meV and 13 meV cannot be separately resolved. Instead, we plot the linewidth (FWHM) of the merged excitations centered at 13 meV as a function of polarization. This quantity can act as an independent indicator for the presence of an additional mode around 12.5 meV. Indeed, the subtle two-fold symmetry of the linewidth, plotted in Fig. 3(i), alludes to the existence of a 12.5 meV shoulder with two-fold symmetry, consistent with the observation made for sample A.

Recent theoretical studies have put forward Pomeranchuk fluctuations of a reconstructed Fermi surface or smectic bond-density-wave fluctuations as SC pairing mechanisms~\cite{lin-21,tazai-21}. Our highly polarization-resolved Raman data at $T=5$ K on both samples evidence a weak twofold anisotropy, which could be a vestige of an emerging symmetry breaking within the superconducting phase. Given that the three CDW orders of $2\times2\times2$, $2\times2\times4$, and $1\times4$ structures are reported, their mutual competition or interplay may cause intriguing electronic instabilities. As the CDW order accompanies lattice distortions of twofold symmetry and the phonon self-energy changes due to an opening of the CDW gap, any viable theory of explaining the CDW orders should take into account electron-phonon coupling.

\section{Conclusion}

In summary, our Raman scattering study of CsV$_3$Sb$_5$ discloses secondary electronic instabilities in the CDW state, which involve lattice distortions and rotational symmetry breaking with the two-fold degenerate E$_{2g}$ phonon mode. The intertwined electronic and lattice instabilities and electron-phonon coupling will pose constraints on future theories of comprehending two-stage CDW transitions and the relationship between CDW and SC.

\begin{acknowledgments}
We thank Inkyung Song and Tae-Yeong Park for important experimental support. This work was supported by the National Research Foundation (NRF) of Korea (Grants No. 2020R1A2C3012367, and No. 2020R1A5A1016518) and by the Institute for Basic Science (Grant No. IBS-R009-Y3). The work at SKKU was supported by the National Research Foundation (NRF) of Korea (Grants No. 2020R1A2C3012367, and No. 2020R1A5A1016518). H.C.L. was supported by National Natural Science Foundation of China (Grant No. 11822412 and 11774423), Ministry of Science and Technology of China (Grant No. 2018YFE0202600 and 2016YFA0300504) and Beijing Natural Science Foundation (Grant No. Z200005).
\end{acknowledgments}

\end{document}